\documentclass[aps,prb,amsmath,amssymb,superscriptaddress]{revtex4-1}
\usepackage[english]{babel}
\usepackage[utf8]{inputenc}
\usepackage{graphicx}
\newcommand{\LAO}{LaAlO$_3$}
\newcommand{\STO}{SrTiO$_3$}
\newcommand{\LAOSTO}{\LAO{}/\STO{}}

\newcommand{\dgC}{$\,^{\circ}\mathrm{C}$}
\newcommand{\SrO}{SrO}

\newcommand{\TiO}{TiO$_2$}

\newcommand{\ie}{i.e.\@}
\newcommand{\NCO}{Ca$_2$Nb$_3$O$_{10}$}
\newcommand{\SRO}{SrRuO$_3$}
\newcommand{\BFO}{BiFeO$_3$}

\newcommand{\emissivity}{0.7}
\newcommand{\tempClean}{570\dgC{}} %500, emissivity 0.7
\newcommand{\tempSTO}{660\dgC{}} %580, emissivity 0.7
\newcommand{\tempSTOAnneal}{770\dgC{}} %680, emissivity 0.7
\newcommand{\tempSrOIV}{770\dgC{}} %680, emissivity 0.7
\newcommand{\tempLAO}{750\dgC{}} %666, emissivity 0.7
\newcommand{\pressure}{$5\times{}10^{-5}$\,mbar}
\newcommand{\pressureBase}{$2\times{}10^{-9}$\,mbar}
\newcommand{\fluenceSTO}{1.4\,J/cm$^2$}
\newcommand{\fluenceLAO}{2.0\,J/cm$^2$}

\newcommand{\SampleA}{A} %fig 1,2,3
\newcommand{\SampleB}{B} %fig 4
\newcommand{\SampleC}{C} %fig 5,6,7
\newcommand{\SampleD}{D} %reference on STO

\newcommand{\SampleBpulses}{600}
\newcommand{\SampleCpulses}{588}
\newcommand{\SampleAnm}{3.3\,nm}
\newcommand{\SampleBnm}{5.7\,nm}
\newcommand{\SampleCnm}{5.5\,nm}

\begin{document}

\title{Growing a \LAOSTO{} heterostructure on \NCO{} nanosheets}

\author{Alexander J.H. van der Torren}
\altaffiliation[New address: ] {ASML, Veldhoven, NL} \affiliation{Huygens - Kamerlingh Onnes  Laboratorium, Leiden University, Niels Bohrweg 2, 2300 RA Leiden,
The Netherlands}
\author{Huiyu Yuan}
\altaffiliation[New address: ] {School of Materials Science and Engineering, Zhengzhou University, Zhengzhou, People's Republic of China}\affiliation{MESA+ Institute for Nanotechnology, University of Twente, PO Box 217, 7500 AE Enschede, The Netherlands}
\author{Zhaoliang Liao}
\altaffiliation[New address: ] {Oak Ridge National Laboratory, Tennessee, US} \affiliation{MESA+ Institute for Nanotechnology, University of Twente, PO Box 217, 7500 AE Enschede, The Netherlands}
\author{Johan E. ten Elshof}
\affiliation{MESA+ Institute for Nanotechnology, University of Twente, PO Box 217, 7500 AE Enschede, The Netherlands}
\author{Gertjan Koster}
\affiliation{MESA+ Institute for Nanotechnology, University of Twente, PO Box 217, 7500 AE Enschede, The Netherlands}
\author{Mark Huijben}
\affiliation{MESA+ Institute for Nanotechnology, University of Twente, PO Box 217, 7500 AE Enschede, The Netherlands}
\author{Guus J. H. M. Rijnders}
\affiliation{MESA+ Institute for Nanotechnology, University of Twente, PO Box 217, 7500 AE Enschede, The Netherlands}
\author{Marcel B. S. Hesselberth}
\affiliation{Huygens - Kamerlingh Onnes  Laboratorium, Leiden University, Niels Bohrweg 2, 2300 RA Leiden, The Netherlands}
\author{Johannes Jobst}
\affiliation{Huygens - Kamerlingh Onnes  Laboratorium, Leiden University,
Niels Bohrweg 2, 2300 RA Leiden, The Netherlands}
\author{Sense van der Molen}
\affiliation{Huygens - Kamerlingh Onnes  Laboratorium, Leiden University, Niels Bohrweg 2, 2300 RA Leiden, The
Netherlands}
\author{Jan Aarts}
\affiliation{Huygens - Kamerlingh Onnes  Laboratorium, Leiden University,
Niels Bohrweg 2, 2300 RA Leiden, The Netherlands}

%
%
%\ead{aarts@physics.leidenuniv.nl}
%
\date{\today}
\begin{abstract}
The two-dimensional electron liquid which forms between the band insulators \LAO{} (LAO) and \STO{} (STO) is a
promising component for oxide electronics, but the requirement of using single crystal \STO{} substrates for the growth limits its
applications in terms of device fabrication. It is therefore important to find ways to deposit these materials on other substrates, preferably Si, or Si-based, in order to facilitate integration with existing technology. Interesting
candidates are micron-sized nanosheets of \NCO{} which can be used as seed layers for perovskite materials on any
substrate. We have used low-energy electron microscopy (LEEM) with \emph{in-situ} pulsed laser deposition to study the
subsequent growth of STO and LAO on such flakes which were deposited on Si. We can follow the morphology and
crystallinity of the layers during growth, as well as fingerprint their electronic properties with angle resolved
reflected electron spectroscopy. We find that STO layers, deposited on the nanosheets, can be made crystalline and
flat; that LAO can be grown in a layer-by-layer fashion; and that the full heterostructure shows the signature of
the formation of a conducting interface.
\end{abstract}

\maketitle

%\todoinAT{Samples used:
%\begin{itemize}
%\item Fig.~\ref{fig:nanosheet}, \ref{fig:STOgrowth}, \ref{fig:STO}, sample~A (s60516A\_nanosheet3), 360~pulses \STO, heating, 5\,uc \LAO{}.
%\item Fig.~\ref{fig:STOheating}, sample~B (s60725A\_nanosheet6), 600~pulses \STO{}, heating
%\item Fig.~\ref{fig:LAOgrowth}, \ref{figN:ARRES}, \ref{fig:LAOsurface}, sample~C (s60704B\_nanosheet7), 588~pulses \STO{}, heating, 5\,uc \LAO{}
%\end{itemize}
%}
\section{Introduction}
Transition metal oxide (TMO) perovskites form an interesting group of materials with a large variety of physical
properties, among which the occurrence of superconductivity, of ferromagnetism, and of ferroelectricity~\cite{rao89}.  Moreover, stacking layers of different such oxides allows for new properties to be made on design. A well known example is the formation of the two-dimensional electron system (2DES) between the band insulators \STO{} (STO) and \LAO{} (LAO)~\cite{ohtomo04,gariglio16}. Typically, these interfaces are fabricated by growing a layer of one material (LAO) on a single crystal substrate of the other (STO). The formation of the conducting interface requires the STO crystal or film to be terminated by its TiO$_2$ layer, on which the AlO$_2$ layer of the LAO can grow. The polar discontinuity between the TiO$_2$ layer with its net zero charge and the AlO$_2$ layer with a charge of -{\it e} ({\it e} the electron charge), together with oxygen defects forming during the deposition of the LAO layer, are two important ingredients in the charge transfer to the interface and the formation of the 2DEG \cite{huang18,chen11}. This is the reason that STO substrates are almost exclusively used in researching the 2DES properties. Since only a thin layer of \STO{} is needed to create the 2DES, integration with other materials could profit from the ability to use different substrates. Growth of STO on Si has been demonstrated~\cite{park10}, and conditions were found which allow epitaxial growth, but it would still be useful to have as few constraints as possible for the choice of substrate. Here we develop that approach and use inorganic nanosheets. Such sheets can be obtained through exfoliation of
materials with a lattice closely matching the lattice parameter $a_0$ of \STO{} ($a_0$ = 0.3905~nm). In our case that is \NCO{}~(CNO; a$_0$ = 0.386~nm), on which the growth of epitaxial \STO{}, \SRO{} and \BFO{} has already been demonstrated~\cite{dral15}. For STO in particular, it was found that STO grows unstrained, due to the flexibility of the CNO nanosheet. This approach is very flexible, as nanosheets can be deposited on virtually any substrate~\cite{dral15,nijland14,shibata08,yuan15}, rendering design freedom in tailoring
device properties. The exfoliated nanosheets have a thickness of only 1.5~nm to
4\,nm\cite{yuan15,tetsuka09,osada10,akatsuka12}, which means that they can also be used in experiments which make use of a backgate. \\
\begin{figure}[tb]
\includegraphics[width=\columnwidth]{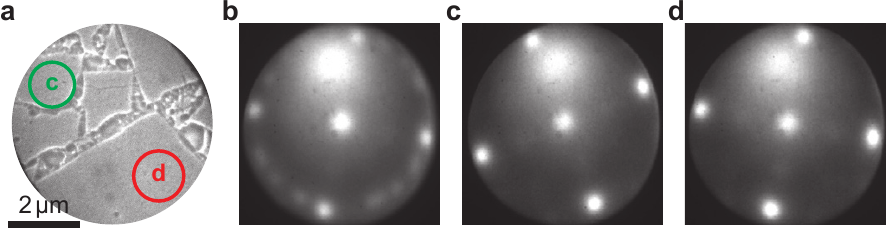}
\caption{ (a) LEEM image of nanosheets on Si (sample~A), taken close to 0\,eV landing energy. (b) The diffraction pattern, taken
at 14\,eV landing energy. Note the numerous weak reflections on the circle between the first order diffraction spots. (c) $\mu-$LEED{}
pattern, also at 14\,eV, recorded from an area indicated with the green circle (upper, left) in (a). (d) the same for the red circle (lower, right). Figure taken from \cite{to-th}.
}
\label{fig:nanosheet}
\end{figure}
To study the growth of LAO/STO on CNO we use low-energy electron microscopy (LEEM) with \emph{in-situ} pulsed laser
deposition (PLD). This combination allows us to follow details of the growth in real time but also to investigate the
electronic structure of the material as we grow it. We recently reported on a LEEM study of the growth of LAO on STO
with various terminations~\cite{torren17}, and we showed that (i) we can observe layer-by-layer growth through the
oscillations in the spot width and intensity of the specular beam in the diffraction pattern as function of time; (ii)
we can image the layer morphology, in particular surface atomic steps; and (iii) most importantly, we can obtain an
electronic fingerprint from the LAO surface during growth, which allows us to determine whether this layer is giving
rise to a conducting or to an insulating interface. The fingerprint is based on angle-resolved reflected-electron
spectroscopy (ARRES), a technique which allows to map the empty electron bands of the layer at energies above the
vacuum energy~\cite{jobst15}. By comparing the results obtained here on nanosheets with our earlier study of \LAOSTO{},
we come to the conclusion that we can grow flat and crystalline heterostructures, and that they show the fingerprints
of a conducting \LAO{}/\STO{} interface~\cite{to-th}.

\section{Results}
\subsection{Growth of the STO template layer}
We discuss four different samples. On samples A we study growth during STO deposition; sample B is used to characterise and improve the surface of the deposited STO layer before depositing LAO; samples C is used to grow 5 unit cells of LAO on top of the STO layer; and sample D is a reference samples where again 5 unit cells of LAO is grown on a TiO$_2$-terminated STO substrate. {\bf Figure~\ref{fig:nanosheet}a} shows a bright field LEEM image of \NCO{} nanosheets deposited on Si (sample~A). Individual nanosheets can clearly be recognized. The square surface net expected for \NCO{} is visible in the LEED pattern (Fig.~\ref{fig:nanosheet}b) taken from the same area as Fig.~\ref{fig:nanosheet}a. Note that the substrate does not contribute LEED spots because of the native oxide on the Si surface. Besides the four principal spots originating from the big flake covering the bottom half of the image, many other rotated patterns are visible originating from the flakes in the top half of the image. This indicates that the nanosheets are randomly oriented on the Si substrate. The rotations of individual flakes can be determined by recording $\mu-$LEED patterns. For example, Figs.~\ref{fig:nanosheet}c,d show that the sheets indicated with green (top) and red (bottom) circles in the LEEM image in Fig.~\ref{fig:nanosheet}a have different orientations. The diffraction spots are not very sharp, due to imperfect
crystallinity or absorbates at the surface\cite{horn99}.

\begin{figure}[tb]
\includegraphics[width=\columnwidth]{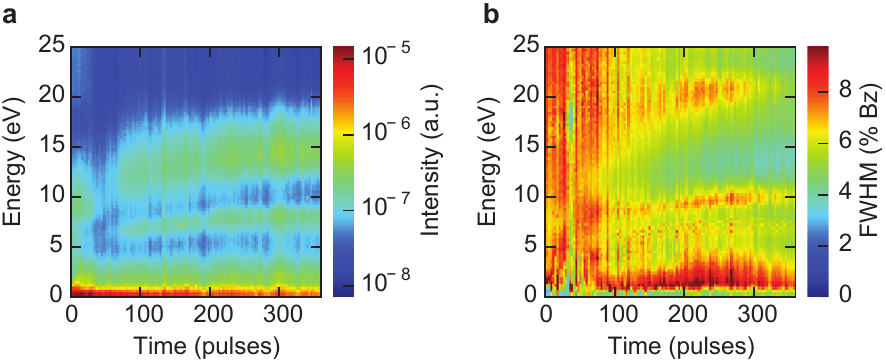}
\caption{
Intensity (a) and FWHM (b) map of the specular (0,0) diffraction spot versus landing energy during deposition of \STO{} on sample~A.
The intensity map in (a)  shows a clear change in material, while the FWHM maps in (b) shows sharpening of the diffraction spots. Moreover,
the FWHM becomes energy dependent. The FWHM is given in percentage of the size of the Brillouin zone. Figure taken from \cite{to-th}.
}
\label{fig:STOgrowth}
\end{figure}

On the sheets we grew STO by PLD at a growth temperature of \tempSTO{}. Higher temperatures showed degradation of the diffraction pattern. In order to follow the evolution of the surface during growth, we recorded IV-curves between 0~eV and 25\,eV after every 5-10 laser pulses. {\bf Figure~\ref{fig:STOgrowth}} shows a color representation of the evolution of the intensity and full-width-half-maximum (FWHM) of the specular diffraction spot (the (0,0) spot) taken from those IV-curves. During growth we observe a clear change of the intensity (Fig.~\ref{fig:STOgrowth}a) and therefore a change in the electronic fingerprint of the surface, which first of all indicates that deposition is taking place.  The IV-curves converge to a constant spectrum after around 300~pulses which consists of maxima around 3~eV, 8~eV and 15\,eV. As we will discuss below, these are spectra typical for \STO{}. The FWHM (Fig.~\ref{fig:STOgrowth}b) is a sensitive measure of the crystallinity of the surface, while disorder yields broadening of the (0,0) spot over all energies. Surface roughness due to the presence of unit-cell size step edges, only influences the FWHM at energies where the incoming and reflected electrons destructively interfere along the out-of-plane axis of the crystal~\cite{horn99}. During growth, the diffraction spots become sharper, quantified by a global reduction of the FWHM (Fig.~\ref{fig:STOgrowth}b) and the FWHM becomes energy-dependent. This leads us to conclude that the crystallinity improves with time, but the surface is still rough.

\begin{figure}[tb]
\includegraphics[width=\columnwidth]{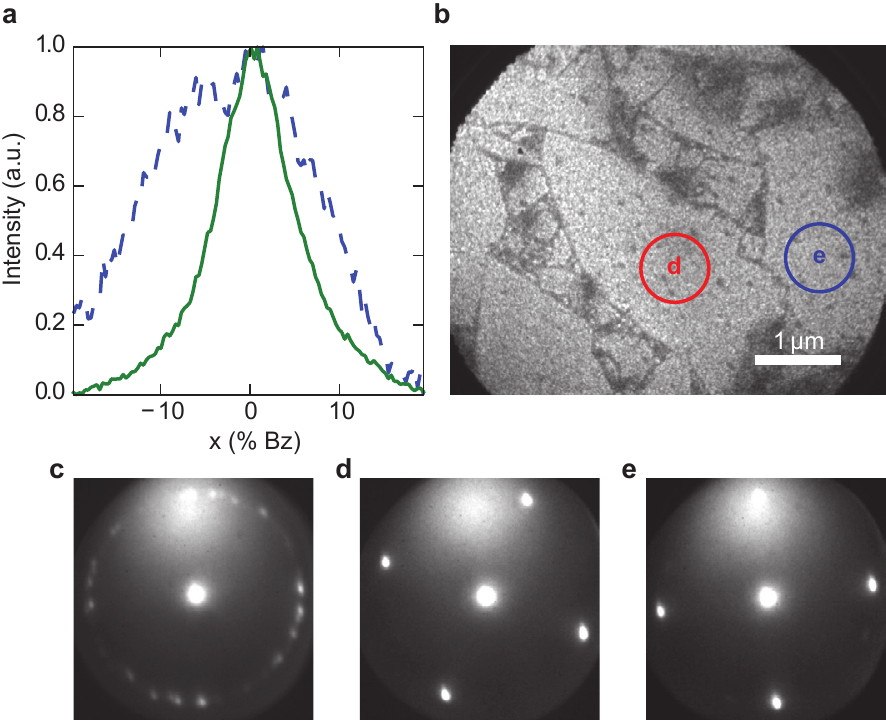}
\caption{
(a) Normalized section of the specular diffraction spot on pure nanosheets (blue, dashed) after annealing, and after growth of \STO{}
(green, solid) on sample~A taken at E$_a$~= 17\,eV. The intensity was normalized, meaning the dashed curve was multiplied by roughly 8 to be visible on the same level as that of the green curve.
(b) Bright field image after growth of 9 u.c. STO at E$_a$~= 29.6\,eV with locations of $\mu-$LEED measurements $d$ and $e$ indicated in red (left) and green (right). (c) LEED pattern of the area shown in b. (d), (e) $\mu-$LEED patterns recorded in the areas indicated in b. The LEED/$\mu-$LEED images in (b,c,d) were taken at $E_a$~=17\,eV. Figure taken from \cite{to-th}.
}
\label{fig:STO}
\end{figure}

%Consequently, the thin film was influenced by the substrate until 300~pulses. From this thickness the surface showed
%bulklike properties, \ie{} more material does not change its band structure further. In other words, this is the minimum \STO{}
%thickness required to form a crystal with complete \STO{} properties.

%During perovskite growth, the film thickness can in most systems be measured up to unit cell precision by monitoring the surface
%roughness, due to the layer-by-layer growth mode these materials grow in. The surface roughness is in most systems monitored by
%reflection high-energy electron diffraction (RHEED). Here instead we used the FWHM of the specular diffraction spot, which has been
%shown to obtain the same result\citeLAOSTO{}.
%However, no oscillations of the surface roughness are observed in figure~\ref{fig:STOgrowth}b. It is not clear whatever the constant large FWHM is
Importantly, neither the intensity nor the FWHM showed growth oscillations, while we observed such oscillations during
the growth of STO and LAO on STO~\cite{torren16,torren17}. For the growth on nanosheets both intensity and FWHM are
constant, and the FWHM is large. It is not clear whether this difference is caused by a small coherence length in the
material or a non layer-by-layer growth mode with a constant rough surface, but it precludes us from an accurate
measurement of the deposition rate. Comparing the growth conditions and amount of pulses to earlier measurements, we
estimate a film thickness close to 3.3~nm or 8.5~unit cells. This means the IV-curves do not change anymore after about
7~unit cells (300 pulses).

{\bf Figure~\ref{fig:STO}a} shows that the deposited layer (presumably STO) grows epitaxially on the nanosheets despite this roughness, from a comparison of a section of the (0,0) spot for the original nanosheets (blue, dashed) and nanosheets with \STO{} (green, solid) taken at E$_a$~= 17\,eV. The curves have been normalized, since the intensity on the nanosheets is 8~times smaller than on the STO surface. The FWHM was reduced by about a factor two in growing STO, indicating a factor 2 increase in the coherent size of the lattice after growth. The individual nanosheets are still clearly visible in the bright field image (Fig.~\ref{fig:STO}b), indicating that the STO only grew epitaxially on the sheets. Indeed, the black areas separating individual nanosheets do not show a diffraction pattern, \ie{} the material there was not crystalline. The LEED pattern in Fig.\ref{fig:STO}c shows multiple diffraction patterns, while {$\mu-$LEED patterns in
Fig.~\ref{fig:STO}d,e show single orientations for every STO/nanosheet heterostructure, demonstrating that the \STO{}
follows the crystal orientation of the underlying nanosheets~\cite{to-th}.

\begin{figure}[tb]
\includegraphics[width=\columnwidth]{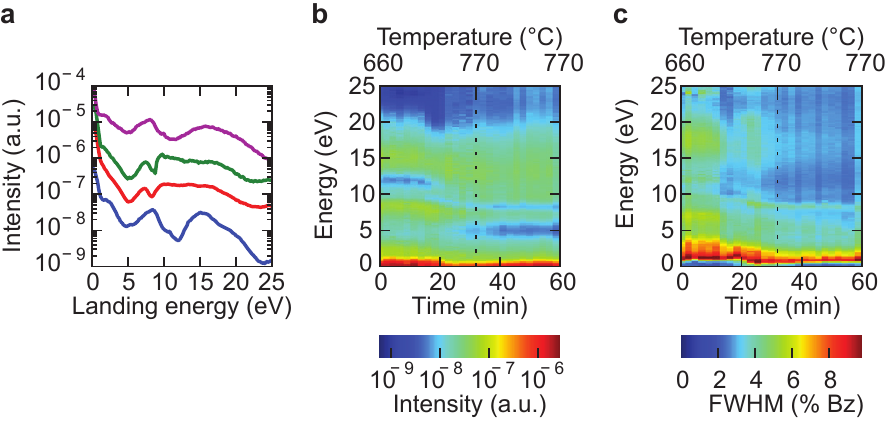}
\caption{
(a) IV-curves before (blue, bottom) and after (red, next curve up) annealing, taken on sample~B. For comparison, two IV-curves taken
on an STO substrate are also shown, one \TiO{}-terminated (green, 2\textsuperscript{nd} from top) and one \SrO{}-terminated (magenta, top).
Curves are shifted for clarity. (b) Intensity and( c) FWHM versus landing energy during the annealing process. The dotted line marks
the time from which the temperature was kept constant after ramping up. The FWHM is given in percentage of the Brillouin zone. Figure taken from \cite{to-th}.
}
\label{fig:STOheating}
\end{figure}
The LEEM-IV spectra can be used to further characterize the deposited layer. {\bf Figure~\ref{fig:STOheating}}a shows IV-curves of the (0,0) spot, taken on sample B (STO thickness of about \SampleBnm{}, \ie{} \SampleBpulses{}~\,PLD pulses). The lowest (blue) curve was taken after deposition, to be compared to a reference curve taken on a \TiO{}-terminated STO substrate (green; 2$^{nd}$ curve from the top) and a curve taken on a \SrO{}-terminated STO substrate (magenta; top curve)~\cite{to-th}. The reference curve was taken from Ref.~\cite{torren17}, while the \SrO{}-termination was obtained by the growth of an \SrO{} bilayer on a \TiO{}-terminated substrate.
The signature of the as-deposited surface clearly shows more resemblance to the \SrO{}-terminated surface, with an intensity minimum around 12~V and a broad maximum around 15~V. In a next step, we heated the samples at constant rate from \tempSTO{} to \tempSTOAnneal{} during 30\,min, and holding the temperature constant for another 30\,min at a constant oxygen pressure of \pressure{} (equal to the pressure used during PLD).

\begin{figure}[t]
\includegraphics[width=\columnwidth]{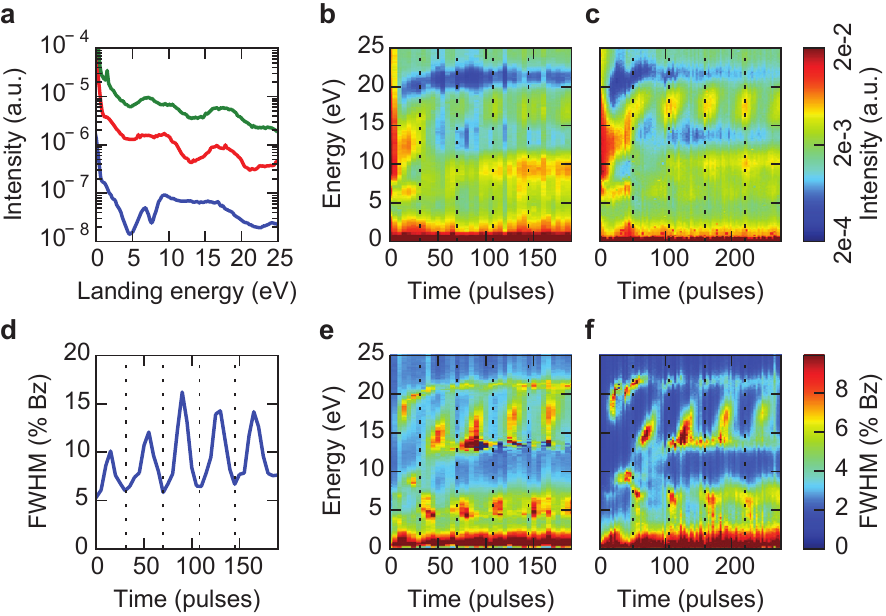}
\caption{
(a) IV-curves before (blue, bottom) and after (red, middle) growth of \LAO{} on sample~C as well as a reference curve of
5~unit cells \LAO{} on a \STO{} substrate on sample~\SampleD{} (green, top). The curves are shifted for clarity. (b) Intensity map versus time and energy of the specular (0,0) diffraction spot on sample~\SampleC{}. (c) the same map on reference sample~\SampleD{}.
(d) FWHM oscillations versus number of laser pulses (time) at E$_a$~= 16\,eV, on sample~\SampleC{}. This energy is close to the out-of-phase conditions where maximal surface sensitivity is reached. (e) FWHM map for all energies on sample~\SampleC{} and (f) FWHM map on sample~\SampleD{}. Figure taken from \cite{to-th}.
}
\label{fig:LAOgrowth}
\end{figure}
The annealing changes the surface electronic properties and also further improves the surface quality. Figure~\ref{fig:STOheating}a shows the IV-curve of the (0,0) spot after annealing (red curve, 2$^{nd}$ from bottom). From the change, and in particular from the appearance of a minimum around 8~eV, we conclude that the surface changes from a dominantly \SrO{}-termination to a \TiO{}-termination. The evolution of the IV curves as function of time is shown in more detail in Fig.~\ref{fig:STOheating}b where it can be seen that there are clearly discernable changes around 20 minutes, that the 8~V minimum has developed around 30~minutes, and that no further changes occur during the next 30~minutes. Figure~\ref{fig:STOheating}c shows the corresponding change of the FWHM of the specular beam. The FWHM clearly decreases in the first 10 - 20 minutes, while the energy dependence disappears, indicating that the surface flattened. Clearly, the change to a \TiO{} termination, the increase in crystallinity, and the flattening of the surface go hand in hand.

\subsection{Growth of LAO}
Next, we use these improved surfaces as growth template for thin films of \LAO{}. In {\bf Fig.~\ref{fig:LAOgrowth}a} we compare the IV-curves before (blue, bottom) and after (red, middle) growth on sample~\SampleC{} (STO thickness of about \SampleCnm{}, \ie{} \SampleCpulses{}\,PLD pulses) with a reference curve of 5~unit cells \LAO{} grown under the same conditions on an STO-substrate (sample~D) (green, top). The IV-curves after growing LAO on the STO/nanosheet sample and after growing on the TiO$_2$-terminated STO substrate are very similar, confirming that similar LAO films were grown in both cases. Moreover, and even more importantly, we observed clear oscillations in both the intensity and the FWHM of the specular beam during the deposition, demonstrating epitaxial growth in layer-by-layer mode. This is analogous to using Reflection High Energy Electron Diffraction (RHEED) in standard PLD growth experiments. The full evolution of the intensity and FWHM maps during growth is shown for sample~\SampleC{} (Fig.~\ref{fig:LAOgrowth}b,c) and for the reference sample~\SampleD{} (Fig.~\ref{fig:LAOgrowth}e,f). The intensity oscillations are particularly clear around E$_a$~= 16\,eV, indicating that both samples grow in the same layer-by-layer fashion. The value of 16~eV is the same as found in earlier experiments, as expected, since it is determined by the energy (or wavelength) where impinging around step edges destructively interfere~\cite{torren17}. As a guide to the eye, vertical dotted lines indicate the full unit cells, showing that 5 unit cells were grown. Growth oscillations are even more clear from the changes in the FWHM at 16\,eV, which are plotted for the nanosheet sample in Fig.~\ref{fig:LAOgrowth}d.
%Comparing the IV-curves to earlier measurements on conducting and non-conducting
%samples in chapter~\ref{chap:LAO-STO} suggests that the interface between the \LAO{} and \STO{} is conducting.
%
\begin{figure}[tb]
\includegraphics[width=12cm]{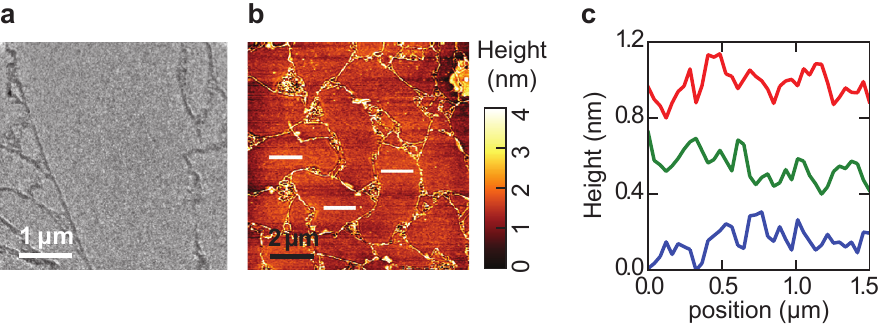}
\caption{
(a) Bright field LEEM image at E$_a$~= 16\,eV and (b) AFM image of the \LAOSTO{}/\NCO{} heterostructure (sample~\SampleC{}). (c) Line profiles of the relative height difference on the nanosheets; profiles from top to bottom correspond in (b) to white bars from left to right. The dark area in the center of the LEEM image is due to an inhomogeneity in channel plate of the detector. Figure taken from \cite{to-th}.
}
\label{fig:LAOsurface}
%Fig.6
\end{figure}
After growth, the shape of the nanosheets is still clearly visible. {\bf Figure~\ref{fig:LAOsurface}} shows a LEEM (a)
and an AFM image (b) taken on sample C. Line profiles taken on a single flake (Fig.~\ref{fig:LAOsurface}c) show a
surface roughness of less than a unit cell (Ra value), pointing to an atomically flat surface. \\

The question remains whether the LAO/STO heterostructure has the desired electronic properties. We studied the LEEM-IV signature of conducting interfaces before~\cite{torren17} and we use those results here. The IV-curves for the LAO/STO/nanosheet sample (Fig.~\ref{fig:LAOgrowth}a) can be characterized as showing a peak around 9~eV, a dip around 13~eV, and another (lower) peak at 17.5~eV. This is the characteristic signature for the conducting LAO/STO interface~\cite{torren17}; for non-conducting interfaces, we find a much broader and higher peak around 18~eV and the suppression of the dip. A more robust distinction can be made by measuring the full angular dependence of the (reflected) intensity using the ARRES technique~\cite{jobst15}. In ARRES we measure electron reflectivity maps not only as a function of landing energy E$_a$ but also of the in-plane wave vector $k_\parallel$ of the electrons. The plot can be represented in the usual Brillouin zone terminology, which in this case spans the high-symmetry points $\Gamma$, $X$ and $M$ of the square surface net.The reflected intensity is determined by the unoccupied states above the vacuum level~\cite{jobst16}. Here, unoccupied bands are presented as intensity minima while band gaps are maxima, because the electrons are reflected with high probability if no states exist in the material at that $E$ and $k_\parallel$. {\bf Fig.~\ref{figN:ARRES}} compares ARRES measurements of a \LAO{}/\allowbreak{}\STO{}/\allowbreak{}nanosheets heterostructure (sample~C) with the reference sample of LAO on a \TiO{}-terminated STO-substrate (sample~D) and maps of an \LAOSTO{} sample with a conducting interface (sample S1-C from Ref.~\cite{torren17}), and a non-conducting interface (sample S2-I from Ref.~\cite{torren17}), respectively. Both maps were taken after deposition of 4 unit cells in a deposition run which continued to 8 unit cells. It should be noted that the data for the nanosheet sample were acquired by taking bright-field real-space images and selecting a single flake in the post processing~\cite{jobst15}, in contrast to the data of Ref.~\cite{torren17} where a 20 $\mu$m$^2$ area was averaged by using LEED data.
\\
%The \LAO{} has been grown under the same conditions and the same thickness.
%
%
\begin{figure}[t]
\begin{center}
\includegraphics[width=\columnwidth]{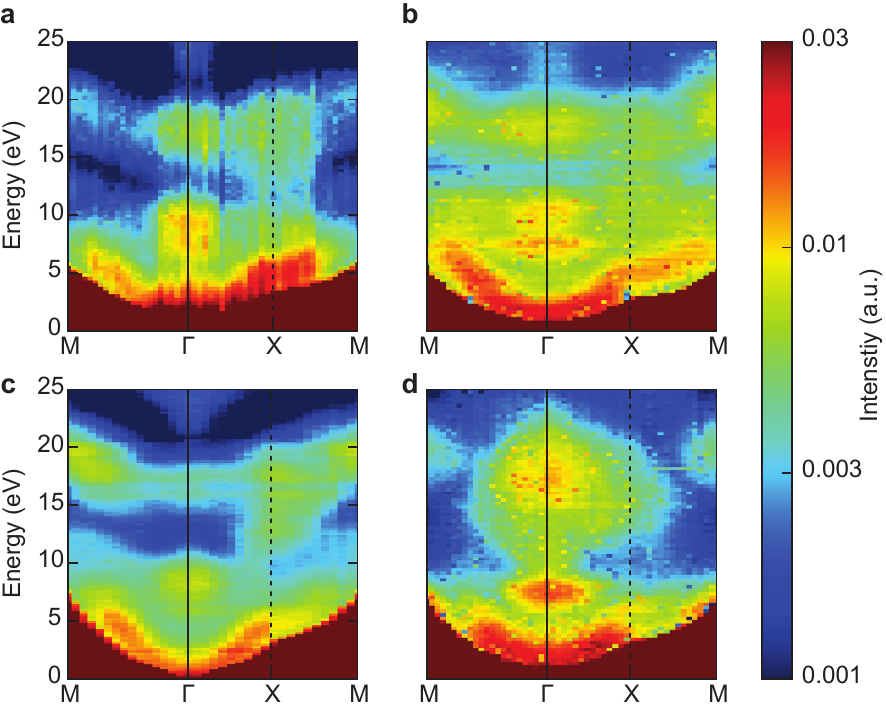}
\caption{ARRES maps from different samples.
(a) \LAOSTO{} on \NCO{} nanosheets (sample~C);  (b) \LAO{} on \STO{} substrate (sample~D). On both samples 5~unit cells of \LAO{} were grown, both images are on the same energy scale. (c) \LAO{} on \STO{} substrate, conducting interface (sample S1-C from Ref.~\cite{torren17}); (d) \LAO{} on \STO{} substrate, non-conducting interface (sample S2-I from Ref.~\cite{torren17}). On both samples 4~unit cells of \LAO{} were grown, both images are on the same energy scale. (c) and (d) were taken from \cite{to-th}.
}
\label{figN:ARRES}
\end{center}
\end{figure}

The salient features for a heterostructure with a conducting interface are a strong peak at the $\Gamma$-point around
9~eV, a weaker peak around 18~eV, and a strong minimum in between; plus high reflected intensity at the $X$-point for
energies between 13~eV and 19~eV. This can be seen in both conducting samples in Fig.~\ref{figN:ARRES}b,c. The intensity distributions are
somewhat different, probably because the first is a 5 u.c. sample and the second a 4 u.c. sample. Around the $\Gamma$-point, the sample grown on the nanosheets (Fig.~\ref{figN:ARRES}a) looks similar to these, and rather different from the insulating sample of Fig.~\ref{figN:ARRES}d, which is characterized by an intensity peak at the $\Gamma$-point around 9~eV, and a roughly equally strong and also very broad peak (in $k$-space) around 18~eV, but little intensity at the $X$-point. Around the $X$-point, the difference between the nanosheet sample and the insulating sample are actually obvious, which is best seen by imagining a line between $\Gamma$ and $X$ at around 17~eV. The conducting samples and the nanosheet sample show a peak-dip-peak intensity variation, while the insulating sample just shows a diminishing intensity. On the whole, the $X$-point differences seem the most important but both from $\Gamma$-point and $X$-point considerations the nanosheet sample shows stronger similarities with the conducting sample S2-C (Fig.~\ref{figN:ARRES}c).
%
%The ARRES signature of the reference sample,~\SampleD{}, actually shows weaker similarities with S2-C, which is mainly
%due to the fact that around the $\Gamma$-point the reflection minimum between 9~eV and 18~eV is not as deep and clear.
%Unfortunately, the sample was lost before the conductance could be measured. Still, we find that the empty band
%structure of the LAO layer is very similar to that of a layer which yields a conducting interface.
\\

%
%This can be solved by selecting the specular diffraction spot during acquisition by an aperture and collecting
%realspace data. A single flake can than be selected during data processing.

%While the ARRES map of \LAO{} on an \STO{}-substrate (Fig.~\ref{figN:ARRES}b) was collected from the specular spot in
%LEED images, averaging electrons from the 5\,\si{\micro}m-wide area covered by the electron beam, this is not possible for
%the \LAOSTO{} heterostructures on the nanosheets due to their lateral inhomogeneity (cf.\ Fig.~\ref{fig:nanosheet}a).
%On the latter, the ARRES map was composed of real-space bright-field images where an area of interest is selected to lie within one nanosheet.

\section{Discussion and conclusions}
The first examples of perovskite growth on \NCO{} nanosheets in the literature already showed they are promising candidates for transfer of TMO devices to literally any substrate. Here we have shown how low-energy electron microscopy studies can help to find the growth conditions and annealing steps required to build TMO devices on \NCO{} nanosheets. In depositing an STO layer directly on top of the nanosheets, we did not observe intensity oscillations in the specular beam as expected for layer-by-layer growth. This is at least partly due to the growth conditions, more specifically the growth temperature. For the first STO layer we could not obtain a crystalline film at temperatures above \tempSTO{} and degradation of the diffraction pattern was observed before the growth started. In the literature however, films are grown at 700~\dgC{}\cite{yuan15,dral15}. The main difference between the conditions in Ref.~\cite{dral15,yuan15} and those reported here is the oxygen background pressure. From this we infer that nanosheets degrade in low oxygen pressures for temperatures above \tempSTO{}. This could be caused by the reduction of oxygen in the nanosheet. On the other hand, the crystallinity of the STO layer clearly improved during growth of the first 7~unit cells and this number seems to be required for a well defined surface with \STO{}-substrate-like properties.
\\
%This is much more than the 1 to 2 unit cells range in which the IV-curve changes when growing \LAO{} on \STO{}.
%
The grown STO film showed a crystalline but rough surface, as can be concluded from the quite broad specular spot . This could be improved by annealing, which again suggests that the growth temperature was too low for ideal growth. The as-grown surface showed an IV-curves characteristic for an \SrO{}-terminated STO sample, while after annealing at \tempSTOAnneal{} the surface showed the IV-curve of
\TiO{}-terminated STO. This suggests that the laser fluence was slightly too low to grow stoichiometric STO. It is well known that the laser fluence influences the Sr/Ti cation ratio. Working with a fluence below the one yielding a 1:1 ratio will result in a film with a (slight) Sr excess~\cite{brecken12}. Growing in this regime of excess Sr can actually lead to a termination conversion when starting on a TiO$_2$-terminated surface, as reported by Bauemer {\it et. al.}~\cite{bauemer15}. Still, that annealing at \tempSrOIV{} changes this is somewhat unexpected since the SrO-terminated reference curve was measured at that same temperature. However, the top layer of the SrO-terminated reference sample was a double layer of SrO forming a stable rock salt structure, and this may be different for the sample we grew here. We surmise that the SrO we observe on the grown SrTiO$_3$ is due to the rough surface and low growth pressure, and unstable at high temperatures. Annealing reorders the surface to a well defined \TiO{}-termination. \\
The roughness, incidentally, also appears to preclude the occurrence of a surface reconstruction. When Ti-deficient growth is started on a flat STO surface, SrO islands form on which a 2x2 reconstruction is found \cite{Xu16}. This surface with half-unit-cell islands on top constitutes a mixed-terminated surface. We do not observe such a surface in Fig.~\ref{fig:STO}b. The only statement we can make is that the surface shows the signature of an SrO termination.
While the nanosheets themselves were unstable at temperatures above \tempSTO{}, no degradation is found during the
annealing of the \STO{} film on nanosheets at \tempSTOAnneal{}. Apparently the STO layer protects the nanosheets from
oxygen reduction. However, the temperature dependence is still fragile and temperatures above \tempSTOAnneal{} are not
recommended at these low pressures. Once a well-defined surface was obtained, layer-by-layer growth of \LAO{} could be
achieved. In the behavior of the intensity oscillations during growth we did not find any difference between the growth of LAO on our STO film on nanosheets and a \TiO{}-terminated STO substrate. The final ARRES map is very similar to ARRES maps found on conducting \LAO{}/\STO{} heterostructures as found in Ref.~\cite{torren17}. From a close comparison with the measurements performed on samples
with conducting and non-conducting interfaces we conclude that the sample grown on the nanosheets shows the characteristics of a conducting interface. This implies not only aTiO2-terminated STO layer, but also the correct LAO stoichiometry. \\
Lastly we stress the opportunity of electronic gating of TMO devices on nanosheets. We infer a minimum \STO{} buffer
layer thickness of only $\approx$ 7~unit cells. For bottom gating of devices this would mean a gate dielectric of only
7~unit cells plus the thickness of the nanosheets, which was reported to be between 1.5~nm and 4\,nm\cite{yuan15,tetsuka09,osada10,akatsuka12}. A gate dielectric of only 7\,nm would provide significant
possibilities for bottom gating. \\

\section{Summary}
We have used a Low Energy Electron Microscope with built-in pulsed-laser-deposition capability to study the feasibility of growing an \STO{}/\LAO{} bilayer with a two-dimensional electron system (2DES) at its interface on lattice-matching nanosheets of \NCO{}, which in turn were deposited on the native oxide on top of a Si substrate. Apart from electron diffraction as a characterization tool, we show how to use the intensity-versus-energy characteristics to obtain more information on the electronic structure of the layers we deposit, and we use Angle-resolved reflected-electron spectroscopy (ARRES) to study the electronic structure of the completed sample. We find conditions leading to a TiO$_2$-terminated STO surface, we demonstrate that we can use intensity oscillations during growth in order to count the number of LAO unit cels, and we infer from ARRES maps that the surface of the sample shows an electronic structure similar to what was found for samples with a conducting interface. The versatility of the nanosheet approach should therefore allow local fabrication of 2DES systems with their own desired geometry. \\

\section{Methods}
\NCO{} nanosheets were synthesized and deposited by Langmuir-Blod\-gett (LB) deposition on Si (001) substrates cut from single crystal wafers with native oxide layers as described elsewhere~\cite{nijland14}. The resulting (unilamellar) layer is about 1~to~2~nm thick. The samples were transferred into the ESCHER LEEM system\cite{schramm11,tromp10,schramm12,tromp13} with in-situ pulsed laser deposition as described in Ref.~\cite{torren16}. PLD growth and LEEM imaging in this system are performed alternatingly, since the high voltage
between objective lens and sample has to be turned off during the sequence of laser pulses. Before starting the growth,
the samples were annealed at \tempClean{} (measured with a pyrometer using an emissivity of \emissivity{}) for at least
half an hour in an oxygen background pressure of \pressure{}, in order to remove any contaminants.

For the deposition of \STO{} onto the nanosheets, the temperature was raised to \tempSTO{}. PLD was performed with a laser fluence of \fluenceSTO{} and a 1\,Hz repetition rate in \pressure{} oxygen~\cite{to-th}. This may result in slightly off-stoichiometric growth of STO \cite{torren16}, but in our microscope it is not possible to increase the pressure and further optimize the growth conditions. Following the deposition, the sample was annealed at \tempSTOAnneal{} for at least half an hour in the same oxygen pressure, in order to improve the surface flatness. The sample was then cooled down and moved to the transfer chamber (\pressureBase{}) while the STO~target was replaced with an LAO~target. After a pre-ablation step of the LAO~target, the sample was transferred back to the measurement position. It was heated to \tempLAO{} at the same oxygen pressure of \pressure{}. \LAO{} was then deposited at a laser fluence of \fluenceLAO{} and 1\,Hz repetition rate. In Ref.~\cite{torren17} it was shown that these parameters result in a conducting LAO/STO interface. Three samples were made with a specific number of laser pulses, namely 360~(sample A), 600~(sample B) and 580~(sample C). The thicknesses we estimate from these pulse numbers, for reference purposes, are \SampleAnm{} (\SampleA{}), \SampleBnm{} (\SampleB) and \SampleCnm{} (\SampleC). As was already mentioned, we estimate thicknesses from comparing the growth conditions and amount of pulses to earlier measurements. In this way we find a film thickness close to 3.3~nm or 8.5~unit cells for sample A (360 pulses).

The nanosheet starting material, the growth process and the final heterostructures were investigated by various techniques available in the LEEM. In particular real-space imaging was used to identify nanosheets, $\mu-$LEED to study crystallinity and roughness, and LEEM-IV spectra (the intensity of reflected electrons as a function of their landing energy E$_a$)~\cite{hannon06} to identify various states of the growing surface. We also used angle-resolved reflected-electron spectroscopy (ARRES)~\cite{jobst15}, which extends the concept of IV-curves by recording electron reflectivity maps as a function of landing energy and in-plane wave vector $k_\parallel$ of the electrons. Such ARRES maps are correlated with the unoccupied band structure of the surface layer and are therefore well suited to probe the
surface stoichiometry of the \LAO{}-layer. This implicitly indicates whether or not a 2-DEG is formed at the \LAOSTO{}
interface as was shown before~\cite{torren17}.

One sample was imaged by Atomic Force Microscopy (AFM), using a standard commercial microscope (Bruker) in tapping mode.

\section*{References}
%\bibliographystyle{unsrt}
%\bibliography{Zotero}
%Uwada, T.; Fuji, S.; Sugiyama, T.; Usman, A.; Miura, A.; Masuhara, H.; Kanaizuka, K.; Haga, M. Glycine Crystallization in Solution by CW %Laser-Induced Microbubble on Gold Thin-film Surface ACS Appl. Mater. Interfaces 2012, 4, 1158- 1163.
%(Note the capitalization of the title is Title Case - first letter of each major word is a capitalized)
%

%

\section*{Acknowledgements}
Part of this work was published as part of the PhD thesis of AvdT~\cite{to-th}. We want to acknowledge Ruud Tromp and Daniel Geelen for discussions and advice and Ruud van Egmond for technical assistance. This work was supported by the Netherlands Organization for Scientific Research (NWO) by means of an ”NWO Groot” grant and by the Leiden-Delft Consortium NanoFront. The work is part of the research programmes NWOnano and DESCO, which are financed by NWO. J.J. acknowledges support from NWO-VENI grant 680-47-447. We would also like to acknowledge networking support by the COST Action MP 1308 (COST TO-BE). \\

\section*{Author Information}
\noindent Corresponding Authors \\
*E-mail: aarts@physics.leidenuniv.nl \\
*E-mail: molen@physics.leidenuniv.nl \\

\section*{Author contribution}
AvdT, JtenE, GK, SvdM and JA contributed to the conception and design of the experiment. HY, ZL, JtenE, GK, GR and MH were involved in the preparation and characterization of the nanolayers. AvdT, MH, and JJ performed the PLD growth, the LEEM measurements and the data analysis. AvdT, JJ, SvdM and JA wrote the manuscript, with assistance of the other authors.

\section*{Additional Information}
{\bf Competing interests:} The authors declare no competing interests.
\end{document}